\documentclass[12pt]{iopart}

\usepackage{iopams}
\usepackage{graphicx,psfrag}
\usepackage{amssymb}
\usepackage{wasysym}
\usepackage{textcomp}
\usepackage{setspace}
\usepackage[dvips,usenames]{color}
\definecolor{violet}{rgb}{0.734,0.558,0.558}
\usepackage{fancyhdr}
\begin{document}

\fancyhf{}
\fancyhead[L]{\textit{\nouppercase{Two-temperature Langevin dynamics}}}
\fancyhead[R]{\nouppercase{V. Dotsenko  \textit{\nouppercase{et al}}}}
\fancyfoot[C]{\thepage}

\title[Two-temperature Langevin dynamics]{Two-temperature Langevin dynamics  in a parabolic potential}

\author{Victor Dotsenko$^{1,2}$, Anna Macio\l ek$^{3,4,5}$,
Oleg Vasilyev$^{3,4}$ and
Gleb Oshanin$^{1}$}

\address{$^1$ Laboratoire de Physique Th{\'e}orique de la Mati{\`e}re
Condens{\'e}e (UMR CNRS 7600), Universit{\'e} Pierre et Marie Curie (Paris 6) -
4 Place Jussieu, 75252 Paris, France}
\address{$^2$
 L.D.\ Landau Institute for Theoretical Physics,
           119334 Moscow, Russia}
\address{$^3$ Max-Planck-Institut f{\"u}r Intelligente Systeme,
  Heisenbergstra\ss e~3, D-70569 Stuttgart, Germany}
  \address{$^4$ Institut f\"ur Theoretische Physik IV,
  Universit\"at Stuttgart, D-70569 Stuttgart, Germany}
  \address{$^5$ Institute of Physical Chemistry, Polish Academy of Sciences,
  Department III, Kasprzaka 44/52, PL-01-224 Warsaw, Poland}

\eads{\mailto{dotsenko@lptmc.jussieu.fr}, 
\mailto{maciolek@fluids.mpi-stuttgart.mpg.de},
\mailto{vasilyev@fluids.mpi-stuttgart.mpg.de},
 \mailto{oshanin@lptmc.jussieu.fr}}

\begin{abstract}
 We study a planar two-temperature  diffusion of a Brownian particle
in  a parabolic potential. The diffusion process is
defined in terms of two Langevin equations with two different
effective temperatures in the $X$ and the $Y$ directions.
In the stationary regime the system is described by
a non-trivial particle position distribution $P(x,y)$, which we determine explicitly.
We show that 
this distribution corresponds to a
non-equilibrium stationary state, characterised
by the presence of space-dependent particle
currents  which  exhibit a non-zero rotor. Theoretical results
are confirmed by the numerical simulations.
\end{abstract} 

\vspace{2pc}


\vspace{2pc}
\submitto{ JSTAT}

\maketitle

\tableofcontents

\section{Introduction}
\label{I}
The idea of physical 
systems characterised by two different temperatures
has been proposed a long time ago for the models 
of spin-glasses and neural networks
with partially annealed disorder
\cite{Penney-Coolen-Sherrington,Dotsenko-Franz-Mezard,Feldman-Dotsenko,Summer-School}.
In these models, the two temperatures $T_{1}$ and $T_{2}$ are related to
two different degrees of freedom, which are evolving at two 
essentially different
time scales. As an example, one may consider a system in which 
the fast spin variables are connected with
the thermal bath kept at the temperature $T_{1}$, 
while the slow spin-spin
coupling parameters are connected with
another thermal bath maintained at the temperature $T_{2}$. 
It can be easily shown that
in the stationary (non-equilibrium) state the statistical properties of
such systems are described by the usual 
replica theory of disordered
systems with a {\it finite} value of the replica parameter $n = T_{1}/T_{2}$
(see also \cite{book1}). Unfortunately, generalisation of this idea to the case
when  dynamics of  two types of degrees of freedom is characterised by
two comparable (or equal) time scales turned out to be rather problematic:
it seems that there is  no generic explicit expression for
the stationary probability distribution function which would generalise the
Gibbs distribution of the equilibrium case $T_{1} = T_{2}$ \cite{Lectures}.
However, there is a particular  case for which one can find
an explicit and a  rather non-trivial expression for the stationary
distribution function. Namely, this is the case when the two degrees of freedom
$x$ and $y$ related to the thermal baths
with the temperatures $T_{x} \not= T_{y}$ respectively, 
experience  a  potential  which is a {\it quadratic} function of 
$x$ and $y$ \cite{Lectures,Ehartier-Peliti}.
During last decade theoretical investigations of such type of systems
were mostly concentrated on  the studies of nonequilibrium fluctuations 
and energy transfer \cite{fluctuations}.
Recently this type of model was studied both theoretically \cite{entropy-memory} and 
experimentally \cite{experiment} 
from the point of view of the entropy production and memory effects.
In this paper, keeping in mind putative experimental realisation of such a type of
systems, we are going to discuss the two-temperatures situation
reformulated in terms of the two-dimensional  diffusion of
a Brownian particle in a parabolic potential.
The diffusion process is defined in terms of  Langevin dynamics
with two different effective temperatures in the $X$ and the $Y$ directions.
In the stationary state this system is described by
a non-trivial  distribution function $P(x,y)$,
which can be computed explicitly. Unlike for the equilibrium case
($T_{x} = T_{y}$), this non-equilibrium stationary state is characterised
by the presence of nontrivial space dependent particle's
flows ${\bf j}(x,y)$. Moreover, these flows exhibit a "symmetry breaking"
rotor,  $S(x,y) = {\boldsymbol \nabla} \times {\bf j}(x,y)$
(directed perpendicular to the $(X,Y)$-plain), the sign (or the direction)
of which is determined by the temperature 
difference $(T_{x}-T_{y})$.

The paper is organised as follows. In Section \ref{model} we define our model
 and present the explicit solution for the stationary
particle's probability distribution function $P(x,y)$.
In Section \ref{obs} we compute putative "observable" quantities of the 
system, such as the variances of the particle displacements  in the $X$ and the $Y$ directions, the
rotor $S(x,y)$ of the particle's flows as well as the average
rotation velocity. In Section \ref{sims} we report the results of the numerical
simulations and compare them with our analytical predictions.
Finally, in Section \ref{conc} we
conclude with a brief recapitulation of our results. 

\section{The model}
\label{model}

We consider  stochastic, over-damped Langevin dynamics of  a  
particle moving 
in a two-dimensional space in a presence of an external potential $U(x,y)$.
The particle instantaneous position $\rho(t)$ is 
defined by  projections on the $X$ and the $Y$ axes, $x(t)$ and  $y(t)$,
respectively. The time evolution of $x(t)$ and $y(t)$ 
is described by following equations:
\begin{eqnarray}
\nonumber
\frac{d}{dt} x(t) &=& -  \frac{\partial}{\partial x} U(x,y)  \; +  \; \xi_{x}(t)
\\
\label{1}
\\
\nonumber
\frac{d}{dt} y(t) &=& -  \frac{\partial}{\partial y} U(x,y)  \; +  \; \xi_{y}(t)
\end{eqnarray}
Here $ \xi_{x,y}(t)$ is  {\it anisotropic} stochastic noise, with zero mean and correlation function 
\begin{equation}
\label{3}
\left<\xi_{\alpha}(t) \xi_{\beta}(t')\right> \; = \; 2 \, T_{\alpha} \, 
\delta_{\alpha,\beta} \, \delta(t - t')
\; , \; \; \; \; \; \; \; (\alpha,\beta \; = x, \; y)
\end{equation}
where $T_{x}$ and $T_{y}$ are two {\it different}  "temperatures"
and $U(x,y)$ has the following parabolic form :
\begin{equation}
\label{4}
U(x,y) \; = \; \frac{1}{2} \,  x^{2} \;  + \; \frac{1}{2} \, y^{2} \; + \; u \,  xy,
\end{equation}
The shape of the potential is controlled by the parameter $u$.
To keep the particle localised near the origin,
we have to impose the constraint $|u| < 1$. This follows from the requirement that both
eigenvalues of the potential, $\lambda_{1,2} = 1 \pm u$, must be positive;
in the case $|u| > 1$, there is a direction in the plane $(x,y)$
at which the potential $U(x,y)$ has a negative
curvature which allows the particle to escape to infinity.

\vspace{5mm}

In the stationary regime,  
the probability distribution function $P(x,y)$ of the particle position  
obeys the stationary Fokker-Planck equation :
\begin{equation}
\label{5}
\fl
\frac{\partial}{\partial x}\Bigl[T_{x}\frac{\partial P(x,y)}{\partial x}  \;  + \;
   P(x,y) \frac{\partial U(x,y)}{\partial x} \Bigr] \; + \;
\frac{\partial}{\partial y}\Bigl[ T_{y}\frac{\partial P(x,y)}{\partial y}  \;  + \;
   P(x,y) \frac{\partial U(x,y)}{\partial y} \Bigr] \; = \; 0
\end{equation}
In the trivial isotropic case, $T_{x} = T_{y} = T$, the solution of the above equation
is simply
 the equilibrium Gibbs distribution $P_{iso}(x,y) \propto \exp\{-\frac{1}{T} \, U(x,y)\}$.

\vspace{5mm}

One can easily show that in the generic {\it anisotropic} case with arbitrary $T_{x}$ and $T_{y}$,
the solution of the stationary equation (\ref{5}) reads:
\begin{equation}
\label{6}
P(x,y) \; = \; Z^{-1} \exp\Bigl\{
-\frac{1}{2} \gamma_{1} \, x^{2}
-\frac{1}{2} \gamma_{2} \, y^{2}
-u\gamma_{3} \, xy
\Bigr\}
\end{equation}
where the following shortenings have been used
\begin{eqnarray}
 \label{7}
\gamma_{1} &=& \frac{T_{x} + \frac{1}{2} u^{2}(T_{x}-T_{y})}{T_{x}T_{y}\bigl(1 + u^{2} \Delta^{2}\bigr)} ,
\\
\nonumber
\\
\label{8}
\gamma_{2} &=& \frac{T_{y} + \frac{1}{2} u^{2}(T_{y}-T_{x})}{T_{x}T_{y}\bigl(1 + u^{2} \Delta^{2}\bigr)} ,
\\
\nonumber
\\
\label{9}
\gamma_{3} &=&  \frac{T_{x} + T_{y}}{2T_{x}T_{y}(1 + u^{2} \Delta^{2})} ,
\end{eqnarray}
and
\begin{equation}
\label{10}
\Delta \; = \;
\frac{(T_{y}-T_{x})}{2 \sqrt{T_{y} T_{x}}} .
\end{equation}
Further on, $Z$ is the normalisation constant (the "partition function"), defined as
\begin{eqnarray}
\nonumber
Z &=& \int\int_{-\infty}^{+\infty} dx \, dy \,
 \exp\Bigl\{
-\frac{1}{2} \gamma_{1} \, x^{2}
-\frac{1}{2} \gamma_{2} \, y^{2}
-u\gamma_{3} \, xy
\Bigr\} \\
\nonumber
\\
&=&
2\pi \, \sqrt{\frac{T_{x}T_{y}\bigl(1 + u^{2} \Delta^{2}\bigr)}{1-u^{2}}} .
\label{11}
\end{eqnarray}
One immediately observes  that $Z$ exists, so that the system has
the stationary solution, only for $|u| < 1$.

\section{The observable quantities}
\label{obs}

\subsection{Variances of particle positions}

Using the above probability distribution function we can straightforwardly calculate the variances of the particles position
with respect to the $X$ and the $Y$ axes:
\begin{eqnarray}
 \label{12}
\langle x^{2}\rangle &=& \frac{T_{x} + \frac{1}{2} u^{2}(T_{y}-T_{x})}{1 - u^{2}} ,
\\
\nonumber
\\
\label{13}
\langle y^{2}\rangle &=& \frac{T_{y} + \frac{1}{2} u^{2}(T_{x}-T_{y})}{1 - u^{2}} .
\end{eqnarray}
The characteristic quantity, which can serve as the measure of anisotropy in  
the  system under study, is defined as the ratio of these two quantities : 
\begin{equation}
\label{14}
g\bigl(T_{y}/T_{x}; \; u\bigr) \; \equiv \;
\frac{\langle x^{2}\rangle}{\langle y^{2}\rangle} \; = \;
\frac{
2 + u^{2}\bigl(T_{y}/T_{x}-1\bigr)}{
2T_{y}/T_{x} + u^{2}\bigl(1-T_{y}/T_{x}\bigr)}
\end{equation}
In the trivial decoupled case, $u=0$, we find
$g\bigl(T_{y}/T_{x}; \; 0\bigr) = T_{x}/T_{y}$,
while in the isotropic case, $T_{x} = T_{y}$ we have
$g\bigl(1; \; u\bigr) = 1$ for all values of the coupling parameter $u$.
Note next that in the strongly anisotropic case, e.g., when  $T_{y}/T_{x} \gg 1$,
one has
\begin{eqnarray}
 \label{15}
\langle x^{2}\rangle &\simeq& \frac{ u^{2}}{2 (1 - u^{2})} T_{y}
\\
\nonumber
\\
\label{16}
\langle y^{2}\rangle & \simeq&  \frac{2- u^{2}}{2 (1 - u^{2})} T_{y}
\\
\nonumber
\\
\label{17}
g &\simeq& \frac{u^{2}}{2 - u^{2}}
\end{eqnarray}
In other words, in the strongly anisotropic case the values of both
$\langle x^{2}\rangle$ and $\langle y^{2}\rangle$ are defined
by the largest $T$, while the value of the ratio
$g = \langle x^{2}\rangle/\langle y^{2}\rangle$ becomes a $T$-independent constant.

\subsection{Mean rotation velocity}

In the stationary case the current ${\bf j} = (j_{x}, j_{y})$
is defined as follows:
\begin{eqnarray}
 \label{18}
j_{x} &=&
T_{x} \frac{\partial P(x,y)}{\partial x}  +    P(x,y) \frac{\partial U(x,y)}{\partial x}
\\
\nonumber
\\
\label{19}
j_{y} &=&
 T_{y}\frac{\partial P(x,y)}{\partial y}  +   P(x,y) \frac{\partial U(x,y)}{\partial y}
\end{eqnarray}
Using eqs.(\ref{4})-(\ref{6}) we obtain :
\begin{eqnarray}
 \label{20}
j_{x} &=&
\Bigl[(1 - T_{x}\gamma_{1}) x + u (1 - T_{x}\gamma_{3}) y  \Bigr] \, P(x,y)
\\
\nonumber
\\
\label{21}
j_{y} &=&
\Bigl[(1 - T_{y}\gamma_{2}) y + u (1 - T_{y}\gamma_{3}) x   \Bigr] \, P(x,y)
\end{eqnarray}
Note that in the isotropic case, $T_{x} = T_{y} = T$, 
we have $\gamma_{1}=\gamma_{2}=\gamma_{3}=1/T$,
so that ${\bf j} \equiv 0$.
In the anisotropic case $T_{x} \not= T_{y}$ 
the above non-trivial
 pattern of currents can be characterised in terms of the rotor:
\begin{equation}
\label{22}
S(x,y) \; \equiv \; {\boldsymbol \nabla} \times {\bf j}(x,y) \; = \;
\frac{\partial}{\partial x} \, j_{y} - \frac{\partial}{\partial y} \, j_{x}
\end{equation}
In general, the rotor $S(x,y)$ is a rather complicated function of two variables $x$ and $y$, 
but it is remarkable that
the function $S(x,y)$ has a non-zero (and very simple) value at the origin at $x=y=0$:
\begin{equation}
\label{23}
S(0) \; = \; u \, \bigl(T_{x} - T_{y}\bigr) \gamma_{3} \, Z^{-1}
\; = \; \frac{u}{4\pi} \;
\frac{T_{x}^{2}-T_{y}^{2}}{T_{x}^{2} T_{y}^{2}} \, 
\sqrt{\frac{T_{x} T_{y} (1-u^{2})}{(1 + u^{2} \Delta^{2})}}
\end{equation}
Note that this quantity changes sign from minus ("left rotation") at $T_{y} > T_{x}$, 
to plus ("right rotation") at   $T_{y} < T_{x}$.

\vspace{5mm}

Due to the presence of a non-zero particle's current rotor, one finds that the mean 
particle's rotation velocity is also non-zero. 
Indeed, for a given value of the particle's
linear velocity ${\bf v}$ located in the point ${\bf r}$ on the two-dimensional plane, its
angular velocity is
\begin{equation}
\label{24}
\omega(t) \; = \; \frac{1}{r^{2}} \, \bigl({\bf v}\times {\bf r}\bigr)
\end{equation}
where $\bigl({\bf v}\times {\bf r}\bigr)$ is the vector product directed along the
$z$-axis. Thus, the mean rotation velocity $\langle \omega\rangle$
in the limit of an infinite observation time can be defined as follows:
\begin{equation}
\label{25}
\langle \omega\rangle \; = \;
\lim_{\tau\to\infty} \,
\frac{1}{\tau} \int_{0}^{\tau} \, dt \; \omega(t)
\end{equation}
Changing averaging over time by  averaging over ensemble  (which will 
be justified in what follows 
by numerical simulations) 
we get:
\begin{equation}
\label{26}
\langle \omega\rangle \; = \;
\int d^{2} {\bf r} \;
\frac{1}{r^{2}} \, \bigl({\bf j}\times {\bf r}\bigr)
\; = \;
\int_{0}^{2\pi} d\phi
\int_{0}^{\infty} dr \;
\bigl(j_{x} \sin\phi - j_{y} \cos\phi \bigr)
\end{equation}
Here the average current ${\bf j}$ is defined in eqs.(\ref{20})-(\ref{21}).
According to eq.(\ref{6}), the probability distribution function $P(r,\phi)$
can be represented as follows
\begin{equation}
\label{27}
P(r,\phi) \; = \;
Z^{-1} \,\exp\bigl\{ -\frac{1}{2} \, r^{2} \, \Psi(\phi) \bigr\}
\end{equation}
where
\begin{equation}
\label{28}
\Psi(\phi) \; = \;
\gamma_{1} \cos^{2}(\phi) \; + \; \gamma_{2} \sin^{2}(\phi) \; + \;
u \, \gamma_{3} \, \sin(2\phi)
\end{equation}
Substituting the explicit expressions for the  components $j_{x}$ and $j_{y}$ of the current,
eqs.(\ref{20})-(\ref{21}), and using eqs.(\ref{7})-(\ref{10}), we get
\begin{equation}
\label{29}
\bigl(j_{x} \sin\phi - j_{y} \cos\phi \bigr) \; = \;
\frac{u \bigl(T_{y} - T_{x}\bigr)}{2 \; Z} \; r   \;
\Psi(\phi)  \; \exp\bigl\{ -\frac{1}{2} \, r^{2} \, \Psi(\phi) \bigr\}
\end{equation}
Substituting eq.(\ref{29}) into eq.(\ref{26}) and performing simple
integrations we obtain
\begin{equation}
\label{30}
\langle \omega\rangle  \; = \;
 u \, \Delta \, \sqrt{\frac{1 - u^{2}}{1 + u^{2} \Delta^{2}}}
\end{equation}
where the parameter $\Delta$ is defined in eq.(\ref{10}).

One can easily prove that the maximal value of the mean angular velocity
is $\langle \omega\rangle_{max} = 1$, and it is
achieved either in the limits $\Delta \to -\infty$ (which corresponds
to $T_{y}\to 0$ for finite $T_{x}$) or in the limit
$\Delta \to +\infty$ (which corresponds
to $T_{x}\to 0$ for a finite $T_{y}$), and the value of the coupling parameter
$u = 1/\sqrt{\Delta} \; \to \; 0$.

\section{Numerical simulations: Brownian dynamics}
\label{sims}

To verify our analytical predictions and 
the underlying assumption 
that the time-average can be replaced by the ensemble average, we perform 
numerical simulations of appropriately discretised 
Langevin equations  eqs.(\ref{1}). Substituting the potential  $U(x,y)=\frac{1}{2}x^{2}+\frac{1}{2}y^{2}+u x y$
into eqs.(\ref{1}) we first write these equations explicitly: 
 \begin{equation}
 \label{eq:delta0}
\begin{array}{l} 
\dot x(t)= 
-x-u y +\xi_{x}(t)  \\
\dot y(t)= 
-y-u x +\xi_{y}(t)\;,  
\end{array}
\end{equation}
where the variances of the thermal noise components are defined by 
$\left<  \xi_{x}^{2} \right> = 2   T_{x}$, 
$\left<  \xi_{y}^{2} \right> = 2   T_{y}$ and 
$\left<  \xi_{x} \xi_{y} \right> =0$.

Discretising Eq. (\ref{eq:delta0}) with a time step $\Delta t$,
we have :
 \begin{equation}
 \label{eq:delta1}
\begin{array}{l} 
 x(t+\Delta t)=  x (t)-\Delta t(x+u y) 
+g_{x}(t) \sqrt{2 T_{x} \Delta t} \\
 y(t+\Delta t)=  y (t)-\Delta t(y+u x)
+g_{y}(t) \sqrt{2 T_{y} \Delta t}\;, 
\end{array}
\end{equation}
where $g_{x}(t)$ and $g_{y}(t)$ are delta-correlated
random numbers with Gaussian distribution of unit half-width,
$ \Delta t \ll 1$,
$ \sqrt{ 2 T_{x,y}  \Delta t} \ll 1$, which are the
  conditions of a smooth motion.
In that case for a free motion of a particle ($U(x,y)=0$)
which starts at the origin ($( x(0)=0,\; y(0)=0)$),
the diffusion coefficients are  $D_{\alpha}=T_{\alpha}$,
$\alpha=x,y$ and the variances of the displacement are given by
$\left< x^{2}(t)\right>=2 T_{x} t$ and
 $\left< y^{2}(t)\right> = 2  T_{y} t$.
In the case of the symmetric potential ($u=0$), one has in  the stationary regime
 $\left< x^{2}\right> =T_{x}$ and $\left< y^{2}\right> =T_{y}$, independently of  $t$.
For asymmetric potential $u \ne 0$, we will compute the mean 
angular  velocity  $\left< \omega \right>$ given in eq.(\ref{25}) 
and the measure of anisotropy 
$g(T_{y}/T_{x},u)=\left<x^2 \right>/\left<y^2 \right>$
that is described by eq.(\ref{14}).

\begin{figure}[h]
\includegraphics[width=0.49\textwidth]{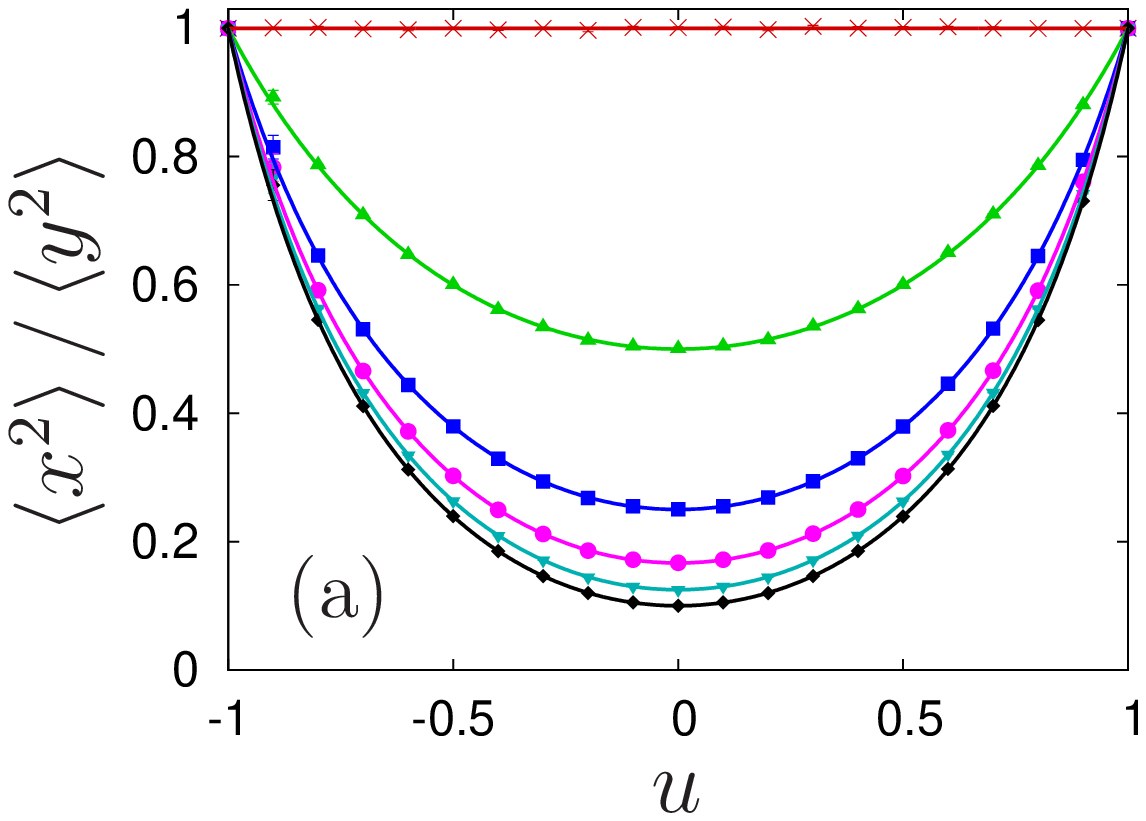}
\includegraphics[width=0.49\textwidth]{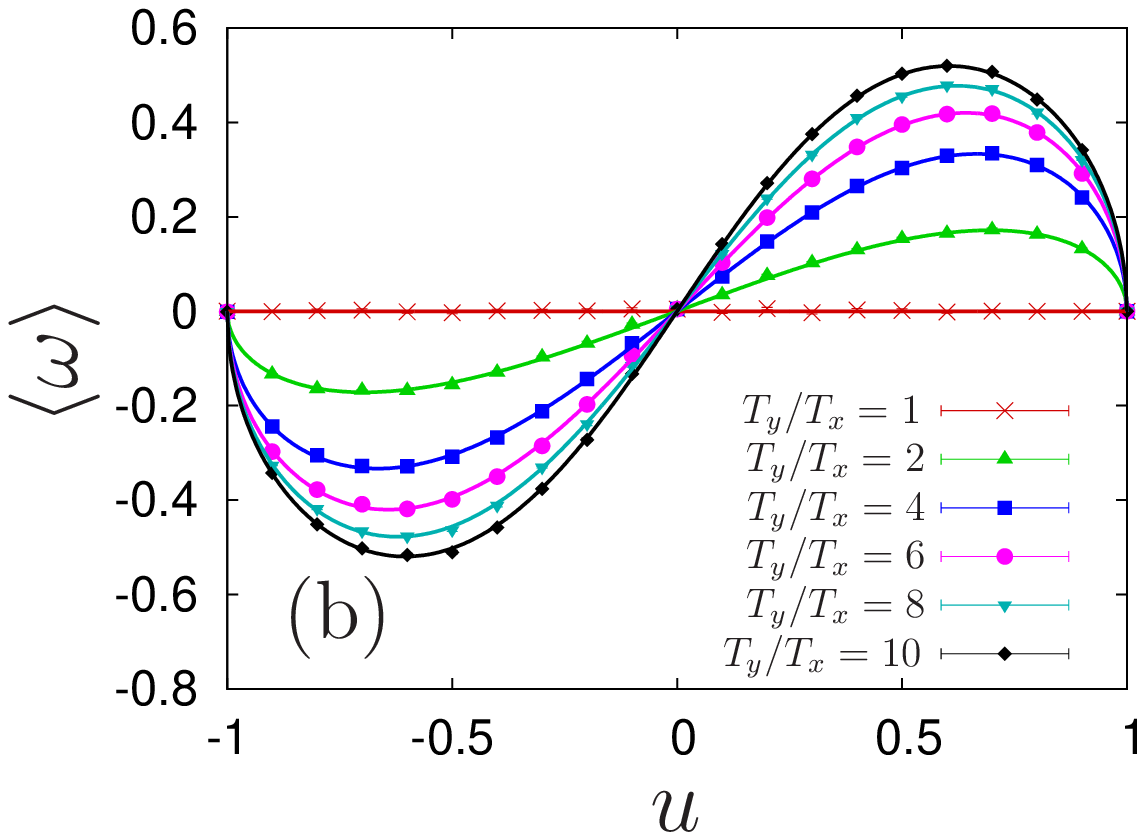}
\caption{
(a) The ratio $\left< x^{2} \right>/\left< y^2\right>$ of  variances of particle's displacements along the $X$ and the $Y$-axes vs the parameter $u$. 
(Symbols  and the color-code is as in (b)~); 
(b) the mean angular velocity $\left< \omega \right>$
 as a function of $u$ for different $T_{y}/T_{x}=1,2,4,6,8,10$. Solid lines are our predictions 
in eqs.(\ref{14}) and (\ref{25}).
 }
\label{fig:u}
\end{figure}

The numerical simulation has been done for the  
time step $\Delta t=0.001$. The averaging has been performed 
over the total time period $\tau=10^{6}$ time units, the numerical inaccuracy has been
evaluated by splitting the whole time interval into $10$ sub-intervals.
In Fig.~\ref{fig:u}(a),(b)  we plot numerical results for
 the ratio of  variances $\left< x^{2}\right>/\left< y^{2}\right>$ 
 and  for the mean angular velocity 
$\left< \omega \right>$, calculated as the time-average of $\omega(t)$, 
as  functions of $u$ for $T_{x}=1$, $T_{y}=1,2,4,6,8,10$. For comparison we also show  
our analytical predictions in
 eq.(\ref{14})
and eq.(\ref{25}), respectively, and find a perfect agreement. This justifies 
 the replacement of the time-average by the ensemble average in our analytical calculations. 

\begin{figure}[h]
\includegraphics[width=0.49\textwidth]{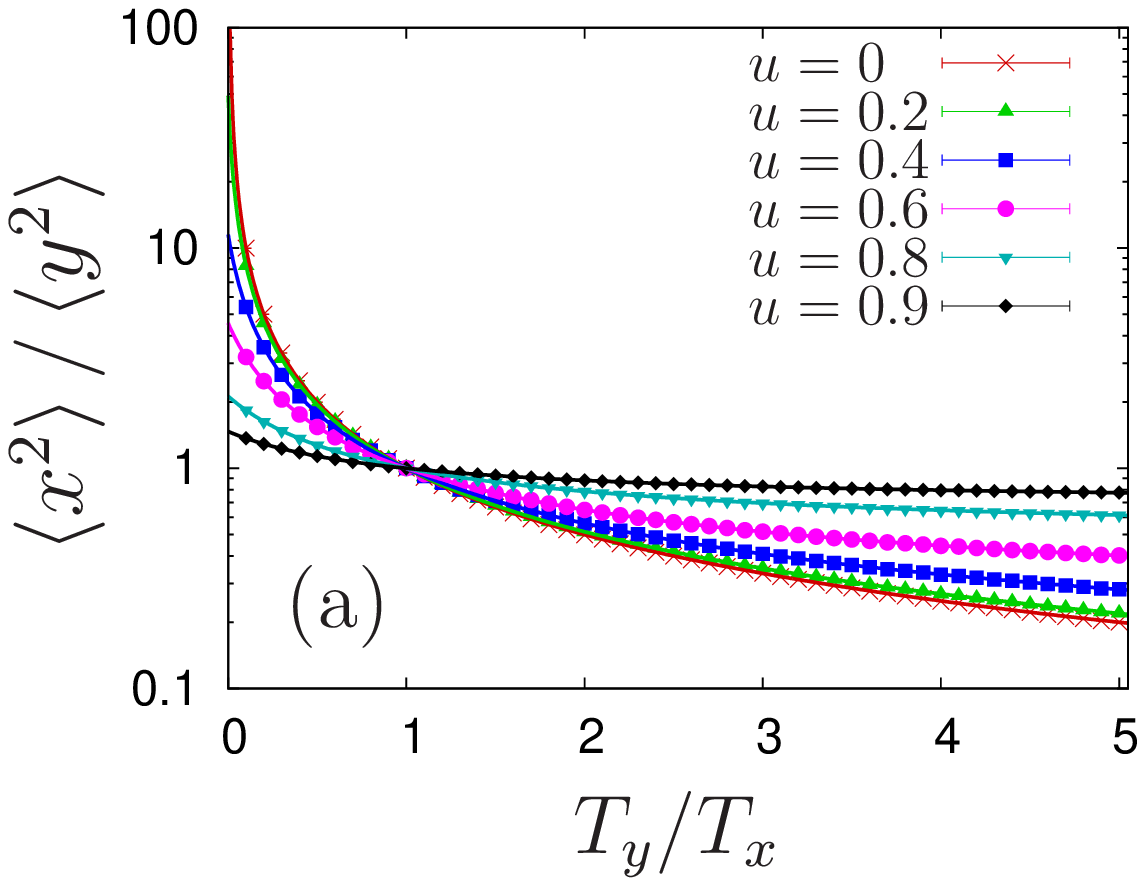}
\includegraphics[width=0.49\textwidth]{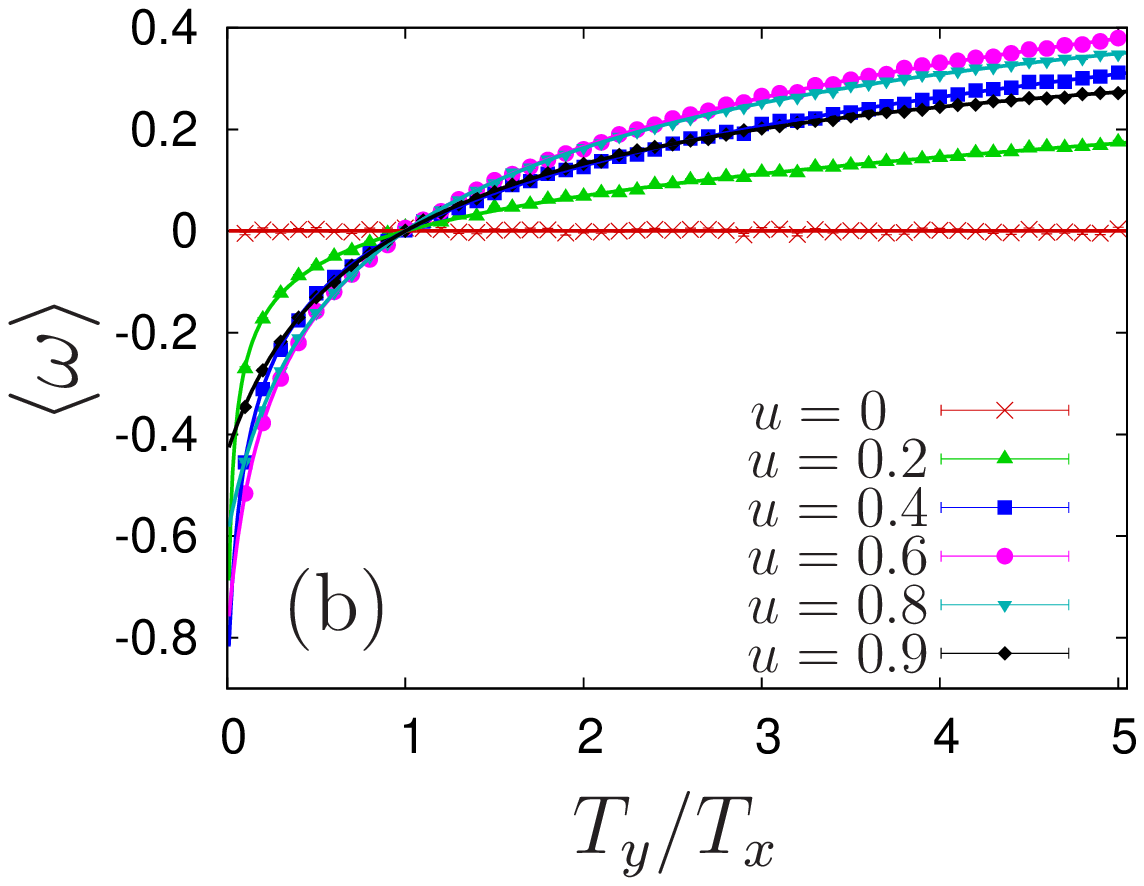}
\caption{
(a) The $\left< x^{2} \right>/\left< y^2\right>$ and 
(b) the mean angular velocity $\left< \omega \right>$ 
 as functions of $T_{y}/T_{x}$ for different values of 
$u=0,0.2,0.4,0.6,0.8,0.9$. Solid lines  
define our theoretical predictions in  eqs.(\ref{14}) and (\ref{25}), and the symbols denote the results of numerical simulations.
 }
\label{fig:t}
\end{figure}

Further on, in Fig.~\ref{fig:t}(a),(b)  we plot  the same quantities
as functions of $T_{y}/T_{x}$ (with $T_{x}=1$) for $u=0,0.2,0.4,0.6,0.8,0.9$.
We again observe a very good agreement between our numerical and analytical results.

\section{Conclusions}
\label{conc}

In the present work we studied a simple stochastic "toy model"
with only two degrees of freedom which are connected to
two thermostats maintained at  two {\it different} temperatures $T_{x}$ and $T_{y}$, respectively.
The model describes diffusion of a particle on a two-dimensional plane
in a presence of a parabolic potential such that the
stochastic noises in the $X$ and the $Y$ directions have different strength
($T_{x}$ and $T_{y}$, respectively).
We determine the stationary state probability distribution function 
for the position of the particle. Despite its relatively simple structure,
 it turns out to be
rather non-trivial, revealing interesting qualitative physical phenomena.
In particular, in the stationary state one finds a rather sophisticated
pattern of  particles' density currents (which would be identically equal to zero
in the equilibrium case) characterised by the
non-zero rotor. Moreover, due to the presence of this flux rotor
one observes the phenomenon which could be interpreted as a
"spontaneous symmetry breaking", namely one finds non-zero value for the
average particle's rotation (around the origin) velocity.
This value is proportional to $(T_{y}-T_{x})$, eq.(\ref{30}),
being positive (left rotation) for $T_{x} < T_{y}$ and
negative (right rotation) for $T_{x} > T_{y}$.

It should be stressed, however, that 
except for recently proposed two-temperature electric analog system
\cite{experiment}, 
for the moment the considered model
has no experimental realization. Thus,
the aim of the present work is somewhat provocative:
we would like argue that the systems of such type are sufficiently
interesting to stimulate investigations for their "hardware" implementations.
We also believe that  modifications of our toy model towards a system that  could be realised 
in practice and at the same time would not loose its interesting behavior (rotation), is possible.

\ack
This work was supported in part by the grant IRSES DCP-PhysBio N269139. 
VD acknowledges the support of Prof. Dietrich at the MPI Stuttgart, where parts of
this work were done.


\section{References}


\begin{thebibliography}{99}

\bibitem{Penney-Coolen-Sherrington} R.W.Penney, T.Coolen and D.Sherrington,
J.Phys. {\bf A26}, 3681 (1993).

\bibitem{Dotsenko-Franz-Mezard} V.S.Dotsenko, S.Franz and M.Mezard,
J.Phys. {\bf A27}, 2351 (1994).

\bibitem{Feldman-Dotsenko} D.E.Feldman and V.S.Dotsenko,
J.Phys. A: Math.Gen., {\bf 27}, 4401 (1994).


\bibitem{Summer-School} V.S.Dotsenko, "Physics of Spin Glasses and Related Problems"
in "The First Landau Institute Summer School, 1993" (Edited by V.Mineev),
Gordon and Breach 1995.

\bibitem{book1} V.S.Dotsenko, "Introduction to the Theory of Spin Glasses and
Neural Networks", World Scientific 1994.

\bibitem{Lectures} V.S.Dotsenko, Lectures on statistical physics of disordered
systems for the Landau Institute graduate students, 1995 (unpublished).

\bibitem{Ehartier-Peliti} R.Exartier and L.Peliti,
      Physics Letters A, {\bf 261}, 94 (1999).

\bibitem{fluctuations} T.Bodineau and B.Derrida, 
     Phys.Rev.Lett. {\bf 92}, 180601 (2004); 
     P.Visco, J.Stat.Mech. P06006 (2006);
     A.Gomez-Martin and J.M.Sancho, Phys.Rev. E {\bf 73}, 045101 (2005)
                       

\bibitem{entropy-memory} A.Crisanti, A.Puglisi and D.Villamaina,
     Phys. Rev. E, {\bf 85}, 061127 (2012); 
                         A.Puglisi and D.Villamaina
     Europhys.Lett. {\bf 88}, 30004 (2009);
                         D.Villamaina, A.Baldassarri, A.Puglisi and A.Vulpiani
     J.Stat.Mech. P07024 (2009)

\bibitem{experiment} S.Ciliberto, A.Imparato, A.Naert and M.Tanase,
     {\it On the heat flux and entropy produced by thermal fluctuations},
     arXiv:1301.4311 (2013)




\end{thebibliography}
\end{document}